\documentclass[2col]{aa}  
\usepackage{graphicx}
\usepackage{txfonts}
\usepackage{natbib}
\usepackage{subfigure}
\newcommand{\der}[0]{\textrm{d}}

\begin{document}
   \title{Non--local radiative transfer in strongly inverted masers.}

   \subtitle{}

   \author{F. Daniel
          \inst{1}
          \and
          J. Cernicharo\inst{1}
          }

   \offprints{F. Daniel}

   \institute{Laboratory of Molecular Astrophysics, Department of Astrophysics, Centro de Astrobiolog\'{\i}a (CSIC-INTA).
              Crta. Torrej\'on a Ajalvir km4. 28840 Torrej\'on de Ardoz. Madrid. Spain\\
              \email{danielf@cab-inta.csic.es}
      }

   \date{}

% \abstract{}{}{}{}{} 
% 5 {} token are mandatory
 
  \abstract
   {Maser transitions are commonly observed in media exhibiting a large range of densities and
   temperatures. They can be 
    used to obtain information on the dynamics and physical 
    conditions of the observed regions. In order to obtain reliable constraints on the physical
    conditions prevailing in the masing regions, it is necessary to model the excitation mechanisms 
   of the energy levels of the observed molecules.}
   {We present a numerical method that enables us to obtain self--consistent 
    solutions for both the statistical equilibrium and radiative transfer equations.}
   {Using the standard maser theory, the method of 
    Short Characteristics is extended to obtain the solution of 
    the integro--differential radiative transfer equation, appropriate 
    to the case of intense masing lines.}
   {We have applied our method to the maser lines of the H$_2$O molecule and
    we compare with the results obtained with a less accurate approach.
    In the regime of 
    large maser opacities we find large differences in the intensity of the maser lines that could be as high as
    several orders of magnitude. The comparison between the two methods 
    shows, however, that the effect on the thermal lines is modest. 
    Finally, the effect introduced by rate coefficients on the prediction of H$_2$O masing lines
    and opacities is discussed, making use of various sets of rate coefficients involving He, o--H$_2$ and p--H$_2$.
%    We find that the masing nature of a line is not affected by the set used but that opacities can vary by up--to a factor
%    $\sim$ 2 from one set to the other.}
    We find that the masing nature of a line is not affected by the selected collisional rates.
    However, from one set to the other the modelled line opacities and intensities can vary by up to a 
    factor $\sim$ 2 and $\sim$10 respectively.}
    {}


   \maketitle
%
%________________________________________________________________

\section{Introduction}

The first maser signal was detected in 1965 \citep{weaver1965}. 
It was due to OH and the intriguing nature of this detection 
was rapidly attributed to a non--thermal phenomenon \citep{davies1967} which
originates in the amplification of the radiation due to stimulated emission. 
Since then, many improvements have been made both 
in observational facilities and maser theory, leading to an abundant literature on the subject.  
The main results, both theoretical and observational, have been summarised by, e.g., 
\citet{elitzur1992} or \citet{gray1999}. From the observational point of view, major 
improvements have been obtained in the last 20 years thanks to the rise of sensitive
interferometers and VLBI techniques \citep[see e.g.][and references therein]{humphreys2007,fish2007}. The sizes of 
the regions where maser emission is detected have typical scales of a few AU.
Nevertheless, some masers arise from large spatial scales, such as the 183.3 GHz line of
water vapour \citep{Cernicharo1990,Cernicharo1994}.
From the point of view of the theory, despite the fact that many improvements have been 
made in explaining the pumping mechanisms of molecular masers, there are no definitive 
explanations that could account for all the observations as yet. 
As an example, the problem of water megamasers observed in the direction of 
AGN is still a subject of debate \citep{elitzur1989,deguchi1989,kylafis1999,Cernicharo2006a}.

The basics of the standard maser theory which is widely used in order to interpret 
masing lines have been reviewed by \citet{elitzur1992}. The main aspect of this theory 
is to introduce the effect of saturation which puts limits on the amount of energy that 
can be radiated by a masing line. As outlined by \citet{collison1995},
maser theoretical studies adopt two different strategies: i) they solve exactly the radiative 
transfer in the maser but with a  simplified description of the energy levels of the molecule;
ii) or deal with an exact description of the molecule but using approximations in the radiative transfer treatment. 
In the latter case, the escape probability formalism is widely used, despite the fact that the validity of this 
approximation is questionable in the case of masing lines. Indeed, intense masers 
require a large amplification path and, therefore, coherence in velocity over relatively large scales. In
contrast, the 
basic principle of the LVG approximation is that the different regions of the cloud are 
decoupled due to Doppler shifting, hence limiting the coherence path. 
In this work we present a self--consistent method that aims to go beyond the approximations usually 
made, i.e. we present a method that enables us to solve exactly the radiative transfer 
problem and in which we deal with a rigorous description of the molecule.     

The theory relevant to the description of masing lines as well as the way 
the radiative transfer is solved is described in Section \ref{theory}.
An application for the o--H$_2$O molecule is presented in Section \ref{application}
where the method developed in this study is compared to 
other less accurate approaches.
Finally, the current method is used in Section \ref{masers_H2O} in order to predict which lines 
of o--H$_2$O or p--H$_2$O could be inverted for densities and temperatures typical
of AGB circumstellar envelopes. In this section, the predictions based on 
various sets of collisional rate coefficients involving He, p--H$_2$ or o--H$_2$ are also compared.

\section{Theory} \label{theory}

\subsection{Absorption coefficient}

In the case of intense masers, the radiation field starts to influence
the inversion of the population, avoiding a complete 
frequency redistribution,
leading to differing line profiles for absorption and emission.
This occurs when the intensity in the masing lines exceeds a certain threshold which depends on the given 
physical conditions. In this case, the maser is said to be saturated.
The way this effect is introduced in the theory is described later in this section.   

Moreover, under a particular geometry
the amplification path varies with direction, implying that the distribution of molecular velocities 
should in principle be considered as anisotropic. This problem was first identified by 
\citet{bettwieser1974}. In the case of spherical geometry, a common approximation, 
known as "standard theory", consists of neglecting the anisotropy of the 
molecular velocities resulting from the existence of directions of 
different path lengths. 
The validity of this assumption was tested in various 
studies \citep[see e.g.][]{bettwieser1976,neufeld1992,emmering1994,elitzur1994} for spherical geometry. 
They concluded that considering explicitly the anisotropies of the 
absorption coefficient could lead to variations
in the properties of the emerging maser radiation, in the case of saturated masers. 
However, it was pointed out that these variations remain unimportant 
in comparison to the uncertainties that affect maser theory, and
that the standard theory should, at least, be a reasonably good first approximation. 
Given the large number of simplifications 
adopted by the standard theory
in the treatment of the radiative transfer (RT), we assume in the following that the masing line absorption 
coefficients are isotropic,
and we follow a procedure similar to that described by \cite{anderson1993}.
First, we consider the frequency--dependent statistical equilibrium 
equations (SEE) for the level $i$:
\begin{eqnarray}
\frac{\der n_i(v)}{\der t} 
& = & \sum_{j\ne i} \left[ N_j \, \phi(v) C_{ji} - n_i(v) C_{ij} \right]
+ \left[ N_i \, \phi(v) - n_i(v) \right] C_{ii} \nonumber \\
& + & \sum_{j \ne i}  \left[ n_j(v) \, R_{ji}(v) - n_i(v) \, R_{ij}(v) \right]
\label{theory:SEE}
\end{eqnarray}
with
\begin{eqnarray}
\begin{array}{cc}
\begin{array}{c}
R_{ij}(v) = 
\end{array}
\left\{
\begin{array}{cccc} \displaystyle
A_{ij} +  \frac{B_{ij}}{4\pi} \int I_v(\Omega)  \, \der \Omega 
& \quad \textrm{if} \quad E_{i} > E_{j} \\ \displaystyle
\frac{B_{ij}}{4\pi} \int I_v(\Omega) \, \der \Omega  & \quad 
\textrm{if} \quad E_{i} < E_{j} \\
\end{array}
\right.
\end{array}
\label{generalite:termes_radiatifs}
\end{eqnarray}
In these equations, $N_j$ represents the total population of level $j$, $n_j(v)$ is
the population of level $j$ at the velocity $v$.
The quantity $\phi(\nu)$ corresponds to the usual Doppler line profile that accounts
 for both the thermal and turbulent motion of the molecules.
$A_{ij}$ and $B_{ij}$ are the
Einstein coefficients for spontaneous and induced emission, $E_j$ is the energy
of level $j$, $C_{ji}$ are the 
collisional rates from level $j$ to level $i$ and $I_v(\Omega)$ is the
specific intensity in a direction specified by $\Omega$. 
We note that in equation \ref{theory:SEE} the
elastic collisional rates $C_{ii}$ have to be formally included since they contribute
to the relaxation of velocities \citep{anderson1993}.
Equation \ref{theory:SEE} can be re-written so that all levels except those 
involving the maser transition are included in the so called pump and loss rates. 
This leads to a phenomenological treatment where the full SEE
is artificially reduced to a two level system.
In the following, the labels $u$ and $l$ indicate the upper and lower maser levels, respectively.
Assuming that for all the other transitions the absorption and emission line profiles
are identical and are given by the Doppler line profile $\phi(\nu)$, 
we obtain from eq. \ref{theory:SEE}:
\begin{eqnarray}
0 & = & \Lambda_u(v) \, \phi(v) - n_u(v) \left[ \Gamma_u(v) + C_{ul} + C_{uu} \right] \nonumber \\
 & + & \phi(v) \left[ N_l \, C_{lu} + N_u \, C_{uu} \right] - n_u(v) \, A_{ul}- \Delta n(v) \, R_{lu}(v) \label {phenom1} \\
0 & = & \Lambda_l(v) \, \phi(v) - n_l(v) \left[ \Gamma_l(v) + C_{lu} + C_{ll} \right] \nonumber \\
 & + & \phi(v) \left[ N_u \, C_{ul} + N_l \, C_{ll} \right] + n_u(v) \, A_{ul} + \Delta n(v) \, R_{lu}(v) 
\label {phenom2}
\end{eqnarray}
where the pump and loss rates, noted $\Lambda_{i}$ and $\Gamma_{i}$ (with $i\in \{ u,l\}$), are defined by
\begin{eqnarray}
\Lambda_{i}(v) & = & \sum_{j \ne \{u,l\}} N_{j} \, \left[ C_{ji} + R_{ji}(v) \right] \\
\Gamma_{i}(v)  & = & \sum_{j \ne \{u,l\}} \left[ C_{ij} + R_{ij}(v) \right] 
\end{eqnarray}
This leads to the usual expression:
\begin{eqnarray}
\Delta n(v) =  \frac{g_l}{g_u} \, n_u(v) - n_l(v)  = n_0(v) \, \left[ 1 + \frac{R_{lu}(v)}{\bar{J}(v)} \right]^{-1} \, \phi(v) \label{difference_pop}
\end{eqnarray}
with 
\begin{eqnarray}
\displaystyle
n_0(v) & = & \frac{1}{1 + \mathcal{B}(v) \frac{g_u}{g_l} A_{ul}} \, \Bigg( \mathcal{A}(v) \left[ \Lambda_u(v) + N_l \, C_{lu} + N_u \, C_{uu} \right] \Bigg. \nonumber \\ 
& - & \Bigg. \mathcal{B}(v) \left[ \Lambda_l(v) + N_u \, C_{ul} + N_l \, C_{ll} \right] \Bigg)\\
\bar{J}(v) & = & \frac{1 + \mathcal{B}(v) \frac{g_u}{g_l} A_{ul}}{\mathcal{A}(v)+\mathcal{B}(v)} \label{theory:int_sat}
\end{eqnarray}
and
\begin{eqnarray}
\mathcal{A}(v) & = & \frac{g_l}{g_u} \left[ \Gamma_u(v) + C_{ul} + C_{uu} + A_{ul}  \right]^{-1}  \\
\mathcal{B}(v) & = & \left[ \Gamma_l(v) + C_{lu} + C_{ll} - \frac{g_u}{g_l} A_{ul}  \right]^{-1} 
\end{eqnarray}
The source function of the masing line is given by:
\begin{eqnarray}
S_{ul}(\nu) = - \frac{2 h \nu^3}{c^2} \, \frac{g_l}{g_u} \, \frac{n_u(\nu)}{\Delta n (\nu)} 
\label{source}
\end{eqnarray}
where $n_u(\nu)$ can be obtained from eqs \ref{phenom1} and \ref{difference_pop}. 
Contrary to the case 
of thermal lines, the source function is frequency dependent 
for the maser line.
 
In order to obtain the source functions 
and the absorption coefficients
from these expressions, we see that the relevant quantities are the specific line intensities (averaged over the angles), 
and the total level populations, $N_j$. 
The latter are obtained by solving the frequency independent SEE expressed by:
\begin{eqnarray}
0
& = & \sum_{j\ne i} \left[ N_j \, C_{ji} - N_i C_{ij} \right] 
+ \sum_{j \ne i}  \left[ N_j \, \bar{J}_{ji} - N_i \, \bar{J}_{ij} \right]
\label{theory:SEE-int}
\end{eqnarray}
with
\begin{eqnarray}
\bar{J}_{kl} = \int \phi_k(\nu) \, R_{kl}(\nu) \, \der \nu 
\end{eqnarray}
In the latter expression, the velocity profile $\phi_k(\nu) = n_k(\nu) / N_k$ accounts for the 
departure 
from the Doppler line profile if the line $k \to l$ is a masing transition, for 
consistency with eq. \ref{theory:SEE}.
In the case of thermal lines, 
these profiles are simply given by the Doppler profile and are identical for both levels. In this latter 
case, the average of the specific intensities over angles and velocities
in eq. \ref{theory:SEE-int} can be factorized, which is crucial in establishing 
the preconditioned 
form of the SEE, as discussed by \citet{rybicki1991}. In the case of masing lines 
and for consistency with eq. \ref{theory:SEE}, 
we have introduced unequal velocity profiles for the maser levels; consequently
the factorisation leading to the preconditioned SEE is no longer feasible.
Hence, the usual ALI method cannot be applied.

In most studies treating the radiative transfer in the presence of masers, 
the pumping rates are taken to be independent of the velocity. In this case
the expression for the population difference is similar to eq.
\ref{difference_pop} except that the velocity dependence of $n_0(\nu)$ 
and $\bar{J}(\nu)$ is omitted. Hence, 
the absorption coefficient will just depend on velocity through the term $R_{lu}(\nu)$.
In the case of unsaturated masers, this term is always negligible in comparison to  
$\bar{J}(\nu)$ and the absorption coefficient is given by the Doppler 
line profile. However, for a saturated maser, i.e., for a line with $R_{lu}(\nu) \sim \bar{J}$, 
the absorption coefficient decreases at the line centre. This translates to the fact 
that, for a given pumping scheme, the molecules pumped to the upper level 
are readily depleted to the lower by stimulated emission since the number 
of masing photons is large. At this stage, the maser amplification starts 
to be linear at line centre and the maser line re--broadens.

\subsection{Solution of the transfer equation}

To solve the transfer equation we have used the short characteristic 
method (SC) introduced by \citet{olson1987} .
The numerical code based on this method has previously been described in 
\citet{daniel2008} for the case of thermal transitions in 1D spherical geometry. 
Briefly, in the SC method we assume 
that the source function can be reproduced numerically by a polynomial function 
of the opacity in the line (usually of order 1 or 2).
At each iteration the absorption coefficients are known from the current 
populations, and the opacities are evaluated assuming that they behave 
linearly between consecutive grid points.
These assumptions permit us to solve analytically the transfer equation 
at each spatial grid point. In our models, the radiative
transfer deals with both maser and thermal lines. 
However, the treatment 
for maser propagation is only switched on when the opacity of a line 
(i.e. $\tau_{ul} = N_l -g_l/g_u \, N_u$), at a particular
grid point, is negative. For the same line and for the radii where the opacity is 
positive, the line is treated in a standard way relevant to thermal radiation.

In what follows, we describe a method that allows us to solve the transfer 
equation for masing lines. It is based on the assumption that the
source functions are linearly interpolated. Hence, along a 
characteristic, specific intensities are derived according to:
\begin{eqnarray}
I_i & = & I_{i-1} \,  \textrm{e}^{-\Delta \tau_i} +
\alpha_i \, S_{i-1} + \beta_i \, S_i
\label{eq2}
\end{eqnarray}
where the $\alpha_i$ and $\beta_i$ coefficients depend on the 
opacity $\Delta \, \tau_i$ between points $i$ and $i-1$.

In the case of masing lines, a difficulty arises from the fact 
that the absorption coefficient at point $i$
depends on the angular average $R_{lu}(v)$ of the specific 
intensities $I_i(\theta)$:
\begin{eqnarray}
\kappa_{lu}(v) = - \frac{h \nu}{4 \, \pi} B_{lu} \, n_0^+(v) \, \left[ 1 + \frac{R_{lu}(v)}{\bar{J}^+(v)} \right]^{-1} \, \phi(v)
\label{theory:coeff_abs}
\end{eqnarray}
where the quantities indexed by a cross (e.g. $\bar{J}^+$) are 
obtained with the current estimate for the populations of the energy levels (i.e.
the populations obtained at the previous iteration).
The same applies to the source functions (see eq. \ref{source}).
Therefore, the radiative transfer equation becomes integro--differential,
whereas in the case of thermal lines, the problem consists 
in solving a differential equation.
Note that the integral part of this integro--differential 
equation involves an integration of the intensities over $4\pi$ steradians. 
In order to obtain intensities which are fully self-consistent with
the level populations used at a given iteration, we thus proceed in three steps.

\subsubsection{First iterative scheme}

As a first step, we derive specific intensities at each grid point 
according to eq. \ref{eq2} assuming that the source functions are 
independent of the radiation field. In other words, we use eq. \ref{source} 
in order to express the source function assuming that the average 
radiation field is $R_{lu}^+(\nu)$.  
For each frequency, we obtain the solution iteratively by calculating 
$I_v(\theta)$ for $\theta \in [0:\pi]$ and then updating $\kappa_{lu}(v)$ 
at grid point $i$. We iterate until the calculated specific intensities 
are consistent with the initial averaged radiation field. 

\begin{figure}
\includegraphics[scale=0.35,angle = 270]{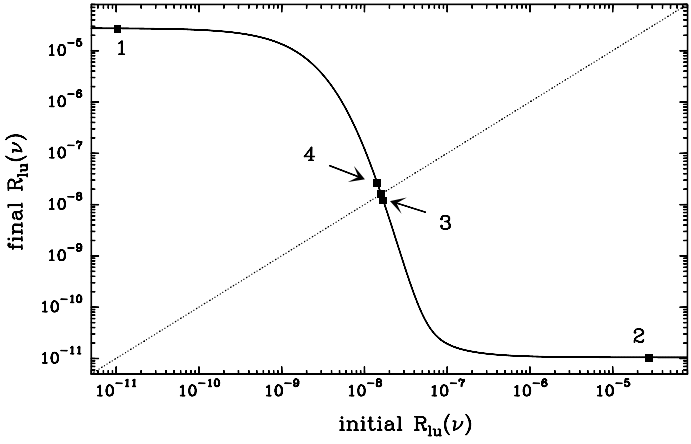}
\caption{Variation of the averaged radiation field (solid line) as a function 
of the initial radiation 
field entering in the determination of the absorption coefficient. The dashed line 
corresponds to a straight line of slope unity. The solution of the problem
corresponds to the intersection between the two curves. The squares correspond
to the averaged radiation field obtained at each iteration.}
\label{theory:sol_trans}
\end{figure}

In the following discussion, $R_{lu}(\nu)$ is indexed by $i$ or $f$ 
whenever it corresponds to the radiation field used in the evaluation 
of $\kappa_{lu}(\nu)$ or to the one which results from 
the evaluation of eq. \ref{eq2} and made use of $R_{lu}^i(\nu)$. 
Additionally, the solutions obtained during the previous steps
are indexed with the superscript
$>$ if they satisfy $R^f_{lu}(\nu) > R^i_{lu}(\nu)$ and with the 
superscript $<$ on the contrary. 
Once the iterative process has been initialised, a trial value 
for $R_{lu}(v)$ is obtained according to:
\begin{eqnarray}
\log R_{lu}(\nu) = \frac{b}{1-a} 
\end{eqnarray}
with
\begin{eqnarray}
 \left \{
\begin{array}{ccc}
a & = & \displaystyle \log \left(\frac{R_{lu}^{f>}(\nu)}{R_{lu}^{f<}(\nu)} \right)
/ \log \left(\frac{R_{lu}^{i>}(\nu)}{R_{lu}^{i<}(\nu)} \right) \\ & & \\
b & = & \displaystyle \log R_{lu}^{f>}(\nu) - a \, \log R_{lu}^{i>}(\nu)
\end{array} \right.
\end{eqnarray}
The radiation field derived in this way is used to update
$\kappa_{lu}(\nu)$. Depending on the
resulting specific intensities, one of the pair
($R_{lu}^{i>}$, $R_{lu}^{f>}$) 
or ($R_{lu}^{i<}$, $R_{lu}^{f<}$) is updated. Finally, the iterative process 
is stopped when we have 
$R_{lu}^{i} = R_{lu}^{f}$, within a given accuracy.
Figure \ref{theory:sol_trans} shows the typical behaviour of $R_{lu}^{f}$ as a function of 
$R_{lu}^{i}$. The averaged radiation fields
determined using our method are 
marked with squares and are indexed according to the iteration number. 

\subsubsection{Second iterative scheme}

Once this process has been achieved, we perform an update of the source 
function at point $i$ according to eq. \ref{source}, and using the 
average radiation field $R_{lu}(\nu)$ derived during the first 
iterative scheme. The whole process is then repeated until we obtain 
convergence for the source function at the current grid point.

\subsubsection{Third iterative scheme}

\begin{figure}
\includegraphics[scale=0.35,angle = 0]{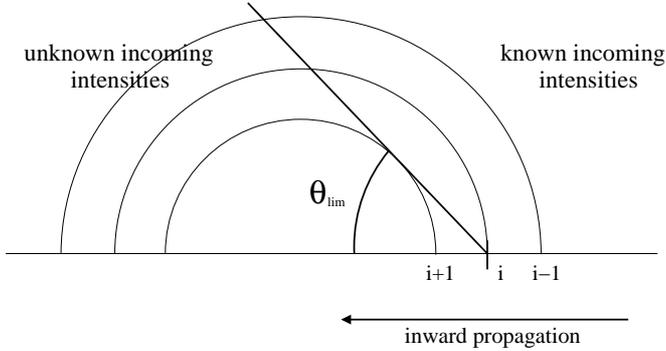}
\caption{
Schematic diagram showing the availability
of incoming 
intensities during the inward
propagation.}
\label{theory:incidente}
\end{figure}

As previously stated, in order to derive the absorption coefficient, it 
is necessary to know the angular average of specific intensities over 
4$\pi$ sr. Using the SC method, the spatial grid is first swept from 
the outermost grid point (i  = N) to the innermost one (i = 1) providing
the intensities $I_i(\theta)$, with $i 
\in [1;N]$ and $\theta \in [\pi/2; \pi])$. The grid is then swept in the 
other direction so that the intensities for $\theta \in [0;\pi/2])$ 
are then known. This sweeping procedure is adopted so that at every moment during the propagation, only the populations used at the current
iteration are needed to derive the intensities.
In the case of masing lines, an additional complexity arises
from the fact that during the inward propagation, the incoming intensities 
for $\theta < \theta_{lim}$ have not yet been calculated 
(c.f. figure \ref{theory:incidente}) with the current populations 
while they enter in the evaluation of the absorption coefficient.

The third iterative scheme thus consists of obtaining an overall convergence 
for the intensities $I_i(\theta)$, with $i \in [1;N]$ and $\theta \in 
[0;\pi]$. This is done by performing successive sweeping of the spatial 
grid, each iteration leading to an update of the intensities $I_i(\theta)$.

\section{Application} \label{application}

In order to test our method, we have performed model calculations
for the ortho state of the water molecule and the comparison
is made with respect to a simplified treatment 
that aims to describe the masing lines. This simplified treatment
relies on the assumption that the calculation of maser inversion 
and the resulting intensities can be achieved in a two--step procedure. First, the 
populations of the molecular levels are determined neglecting the  
specificity of maser radiative transfer. Secondly, the intensities 
of the masing lines are calculated using the populations obtained 
during the first step and taking into account maser propagation.
This procedure has been used in previous studies \citep[e.g.][]{yates1997,watson2002}
and the way it is presently implemented is described in the next sections.

The models consist of clouds of uniform H$_2$ density, temperature (assumed
to be 400 K in all models), and o--H$_2$O abundance (40 levels have been
considered).
We have adopted a diameter for the cloud of
$4.10^{14}$ cm ($2.6\,10^{-4}$ pc) which is the typical size
of bright maser spots in the 22 GHz line of water. 
Moreover, this size provides
a maximum amplification path similar to that of the 
slab depths used in \citet{yates1997}. 
The collisional rate coefficients are for the collisional system 
H$_2$O / He \citep{green1993} and are corrected in order to account 
for the different reduced masses of the H$_2$O / H$_2$ and H$_2$O / He systems. We have
selected these rates rather than those of \citet{dubernet2009} or 
\citet{faure2007} as they were the ones used by \citet{yates1997}. 
These parameters being fixed, we performed the calculations in the 
(n(H$_2$),X(H$_2$O)) plane of parameter space, with respective values in the range 
$[2\,10^5; 8\,10^9]$ cm$^{-3}$ and $[2\,10^{-6} ; 8\,10^{-4}]$.
The accuracy of the results will depend on the gridding adopted. 
Typically, convergence is reached when the 
spatial sampling is such that the opacity between two consecutive grid 
points is of the order or less than 1. For all the calculations 
presented here, we adopted a grid with 500 spatial grid points, 
which might be inadequate for the highest densities/abundances 
considered here. However, since we are interested in masing 
lines which are collisionally quenched at these densities/abundances,
the effect of limited convergence accuracy does not affect our conclusions.
Moreover, the main goal is to compare the effect of the radiative 
transfer and both model calculations are performed with the same spatial grid.

As presented in Section \ref{theory},
elastic collisional rate coefficients for the levels involved 
in masing lines enter in the definition of the absorption coefficients 
and source functions. Test calculations have been performed by 
\cite{anderson1993} in order to test the influence of 
these rates and they concluded that in most cases their
inclusion produces rather small effects. Hence, in the present test calculations,
we have assumed that the elastic rate coefficients are set to zero. In other words, 
in the following calculations we assume that $C_{ii} = 0$ in Eq. \ref{theory:SEE-int, which 
subsequently alter the definition of the pump and loss rates. However, this is not a pre--requisite of the method 
and the elastic rate coefficients could be taken into account, if known. Moreover, despite of this, 
the effect of velocity redistribution is still present through the inelastic collisions terms and through the radiative terms 
associated to transitions which are connected either to the upper or lower energy level 
of the masing line under consideration.}  

\subsection{Pumping scheme with/without maser propagation}

As a first step, we compare the results of the exact calculation 
to a simplified treatment that consists of neglecting the amplification
inherent to maser propagation when deriving the population 
of the energy levels of the molecule.
To do so, we take the absolute value of the 
level population difference which then enters in the definition of 
the absorption coefficients and source functions. These quantities are
defined with the usual expressions which are suitable for thermal lines.
This permit us to treat masing lines in the same manner as thermal lines. This approach was 
previously adopted in plane parallel non--local calculations by 
\citet{yates1997} with the aim of predicting which water lines 
could show maser excitation.
Their results are discussed in term 
of the gain in the masing lines \citep[see eq. 3 of][]{yates1997}
\begin{eqnarray}
\gamma_{ul}(\nu) = \frac{\Delta n(\nu) A_{ul} c^2}{8\pi\nu^2}
\end{eqnarray}
which is subsequently averaged over the slab depths \citep[the so--called depth--weighted 
gain coefficient defined by eq. 7 of ][]{yates1997}.
Since $\gamma_{ul}(\nu)$ is proportional to the absorption 
coefficient $\kappa_{lu}(\nu)$, it is equivalent to consider either the depth--weighted
gain coefficient or the opacity in the masing line, and
in the following, we choose to discuss the results in terms of line opacities.
We emphasise the fact that at this stage, the approximate method described here
does not involve any treatment specific to maser radiation and that could account 
for saturation effects. Thus, the dependence on frequency of the population 
difference $\Delta n(\nu)$ which enters in the definition of 
$\gamma_{ul}(\nu)$ [or $\kappa_{lu}(\nu)$] is just given by the Doppler line profile 
$\phi(\nu)$.

\begin{figure}[h!]
\includegraphics[scale=0.50,angle = 0]{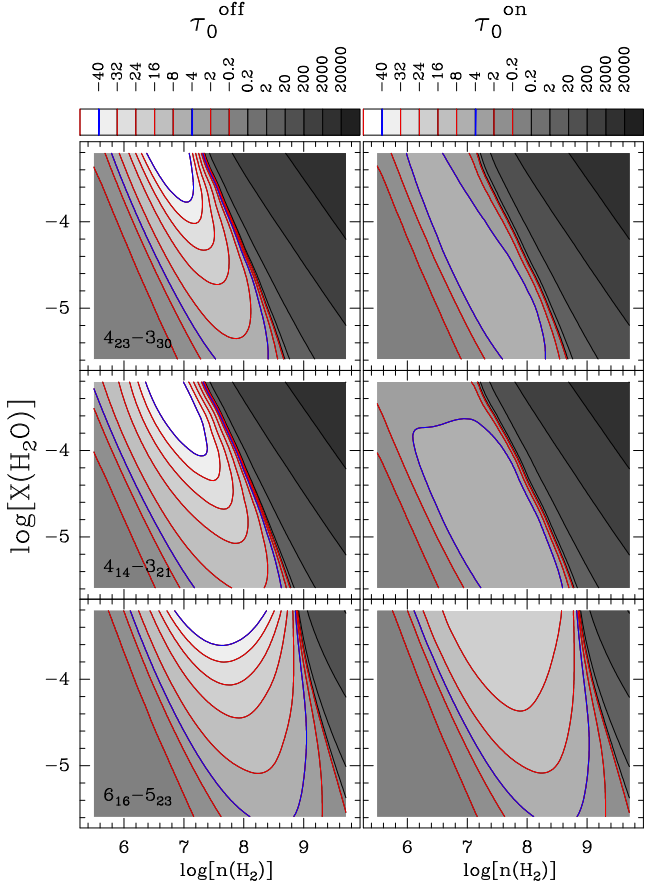}
\caption{
Isocontours of the o--H$_2$O opacities at line centre for the $6_{16}-5_{23}$ line at 22 GHz, 
the $4_{14}-3_{21}$ line at 380 GHz and the $4_{23}-3_{30}$ line at 448 GHz lines.
The left column corresponds to the models where the treatment of 
maser propagation is omitted and the right one to the current 
results (see text). The isocontour values as well as the 
correspondence between grey scale and opacity values are 
displayed on top of the two columns. The isocontours that 
correspond to opacities of -4 and -40 are displayed in 
thick blue lines. Negative values for the opacity correspond 
to red lines.}
\label{app:maser}
\end{figure}

\begin{figure}[h!]
\includegraphics[scale=0.50,angle = 0]{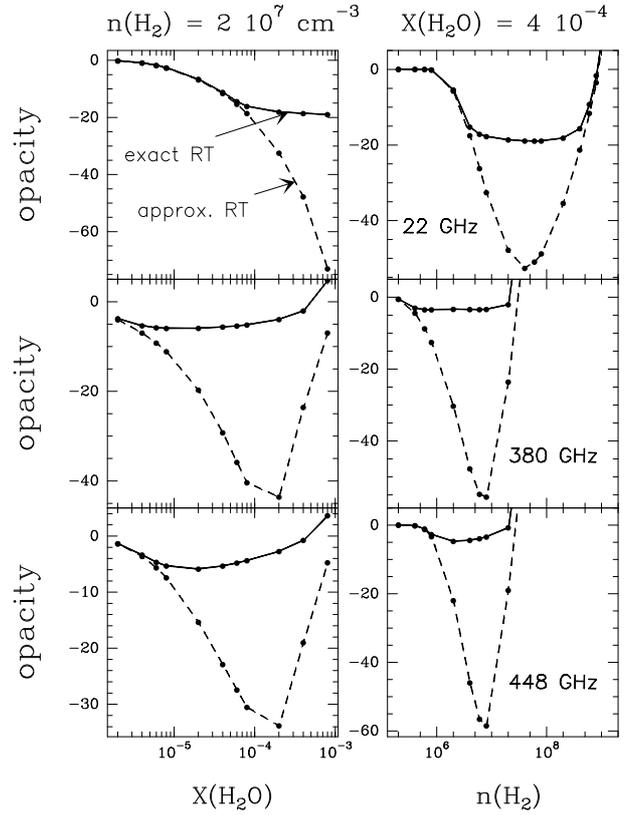}
\caption{
Comparison between the opacities derived with 
(solid lines) or without (dashed lines) accounting for maser 
propagation for the masers at 22 GHz, 380 GHz and 448 GHz. 
The left column corresponds to a cut in density 
at n(H$_2$) = 2 10$^7$ cm$^{-3}$ and the differences are 
shown with respect to variations in X(H$_2$O). The right 
column corresponds to a fixed water abundance, i.e. X(H$_2$O) = 4 10$^{-4}$.}
\label{app:cut}
\end{figure}

Figure \ref{app:maser} shows the opacity at line centre 
of the 22 GHz (6$_{16}$-5$_{23}$), 380 GHz (4$_{14}$-3$_{21}$) and
448 GHz (4$_{23}$-3$_{30}$) masing lines, obtained in the 
two treatments (labelled respectively \textit{on} or 
\textit{off} according to whether maser propagation 
is accounted for or not). Figure \ref{app:cut} shows the 
line opacities for the same lines with respect to cuts 
in density or H$_2$O abundance.
We can see on these figures 
that including the current maser propagation treatment 
leads to a substantial decrease in the predicted line
opacities for these lines. 
Indeed, without treating the exact radiation field in the masing lines,
we obtain minimum optical depths of the order of -70 for 
the 22 GHz line and -40 for the 380 and 448 GHz transitions.
With our method, the minimum optical depth for the 22 GHz line 
is reduced to -20 and to -5 for the other two lines.

\begin{figure}
\includegraphics[scale=0.33,angle = 270]{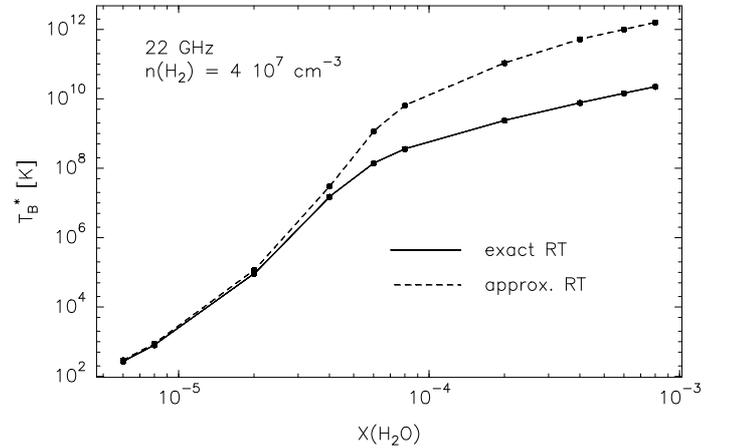}
\caption{Brightness temperatures for the 22 GHz line obtained with the exact treatment (solid lines)
and with the approximated treatment (dashed lines). The intensities correspond to a line of sight that
crosses the centre of the sphere and are given for the rest frequency of the line.}
\label{app:bright22}
\end{figure}

\begin{figure}
\centering
\includegraphics[scale=0.36,angle = 0]{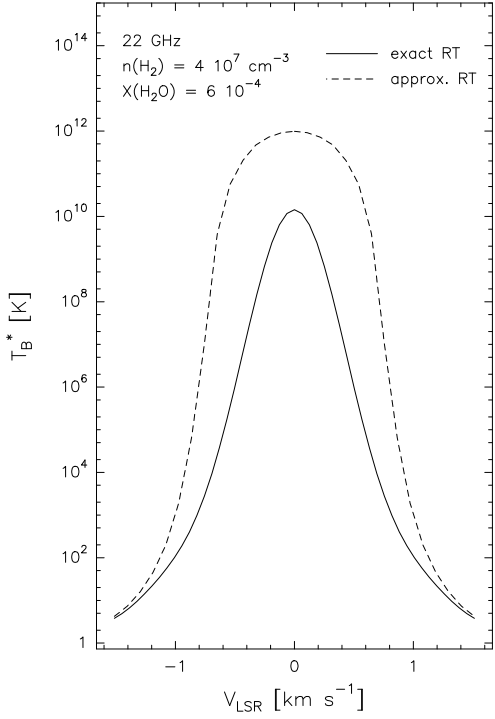}
\caption{Emerging brightness temperature of the 22 GHz line for the 
model with n(H$_2$) = 4 $10^{7}$ and X(H$_2$O) = 6 $10^{-4}$. The solid line 
corresponds to the intensity predicted by the exact calculation and the dashed line
to the one predicted by the approximated RT (see text).
}
\label{app:prof22}
\end{figure}

\begin{figure}
\subfigure[exact treatment]{\includegraphics[scale=0.36,angle = 270]{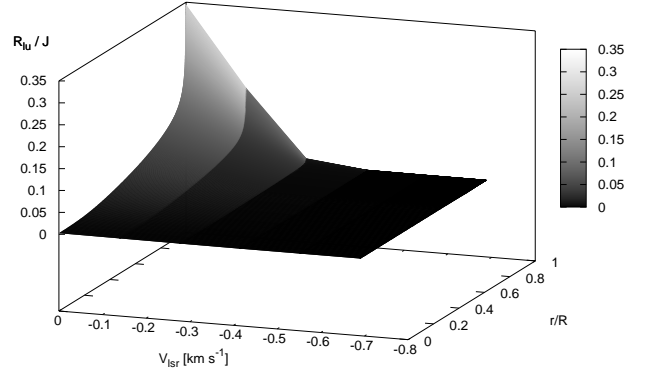}}
\subfigure[approximated treatment]{\includegraphics[scale=0.36,angle = 270]{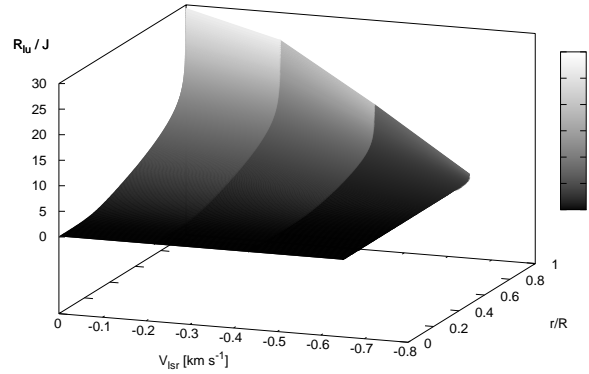}}
\caption{Ratio $R_{lu}(\nu) / \bar{J}(\nu)$ as a function 
of the offset to line centre and of the normalized radius 
in the cloud. This ratio is given for the 22 GHz line 
and corresponds to the model with n(H$_2$) = 4 $10^{7}$ and X(H$_2$O) = 6 $10^{-4}$.
The two plots correspond to (a) the exact treatment and (b) 
an approximated treatment (see text).
}
\label{app:satur22}
\end{figure}

\subsection{Two--level maser propagation}

In order to infer the influence of these changes 
in line opacities, we now consider the resulting masing line intensities.
To do so, any reasonable approximated treatment should include a RT 
theory that accounts for saturation effects.
To obtain a reference point for the current results, we 
%thus 
proceed as follows:
from the populations previously obtained in the approximate treatment, we compute 
the pump and loss rates $\Lambda_i(v)$ and $\Gamma_i(v)$. These 
values enter in the definition 
of the population difference $\Delta n(v)$ through eq. \ref{difference_pop}. Then, for a given masing line,
the intensities are computed by adjusting $R_{lu}(v)$ as described in section \ref{theory}.

Using this approximation, we obtain line intensities that differ largely
from 
%
%the one 
%
those
obtained with the exact treatment described in section \ref{theory}.
This is illustrated in Fig. \ref{app:bright22} for the 22 GHz line. This figure 
shows the brightness temperatures obtained in the two
treatments at a density n(H$_2$) = $4\,10^7$ cm$^{-3}$, 
%for the rest frequency of 
for the centre of
the line and for a line of sight that passes through the centre of the sphere.
%
%This figure shows that the approximate treatment lead to higher intensities that
%differ of 2 orders of magnitude for water abundances X(H$_2$O) $> 5\,10^{-5}$.
%
This figure shows that for water abundances X(H$_2$O) $> 5\,10^{-5}$
the approximate treatment leads to intensities larger by two orders of magnitude.
This originates from the opacity overestimation obtained in the approximate 
treatment which has been discussed in the previous section.
Additionally, Fig. \ref{app:prof22} compares the line profiles computed
in the two treatments for the model with n(H$_2$) = $4\,10^7$ cm$^{-3}$ 
and X(H$_2$O) = 6 $10^{-4}$.
In this figure, we see that the higher intensity found in the 
approximate treatment is accompanied by a broader line profile.
This result is produced by the higher line saturation obtained in the 
approximate treatment with respect to the exact treatment.  
This fact is illustrated in figure \ref{app:satur22} where 
the ratio $R_{lu}(\nu)/\bar{J}(\nu)$ is plotted as a 
function of the velocity offset to line centre and of 
the normalised radius in the cloud for the 22 GHz line and for 
the model with n(H$_2$) = $4 \, 10^7$ cm$^{-3}$ and 
X(H$_2$O) = $ 6 \, 10^{-4}$, for both the exact and 
approximated methods. We see in this figure that in the exact 
calculation the degree of saturation is low, 
i.e. $R_{lu}(\nu)/\bar{J}(\nu) < 1$ at any radii. In contrast, 
the saturation is high in the approximate calculation with $R_{lu}(\nu)/\bar{J}(\nu) > 10$
at most radii.

In the exact treatment, since 
the masing line intensities are taken into account when deriving 
the populations, the population difference between the upper and lower level of 
the masing line tends to be reduced by induced radiative de--excitation.
On the other hand, in the approximate treatment, 
the averaged radiation field is largely 
underestimated. This entails that the amount of induced 
radiative de--excitation events is underestimated,
which leads to a greater difference between the populations of the lower and upper levels.  
Consequently, the derived masing line opacity is larger which 
is accompanied by a larger degree of saturation of the line.
So, the differences encountered in the two treatments originate 
in the overall flow of population from the upper to lower level, which is 
induced by radiative de-excitation. The evaluation of this 
flow of population differs in the two treatments since it 
relies on the accuracy of the determination of the 
intensities in the masing line.  

We note that in section \ref{theory}, eq. \ref{difference_pop} 
is derived from the velocity dependent SEE and from this 
expression it is seen that an increase in the 
average radiation field leads to a decrease in $\Delta n(\nu)$. 
From similar considerations dealing with the velocity 
independent SEE, it can be seen that increasing the value 
of the radiation field for the maser transitions would 
lead to lower population differences for the levels involved 
in the maser transitions. This behaviour is not a specific
case and we can infer from it that considering explicitly 
maser propagation will necessarily lead to a decrease in 
line optical depths in comparison to treatments where the 
average radiation field would be underestimated.

\subsection{Thermal lines}

\begin{figure}
\includegraphics[scale=0.50,angle = 0]{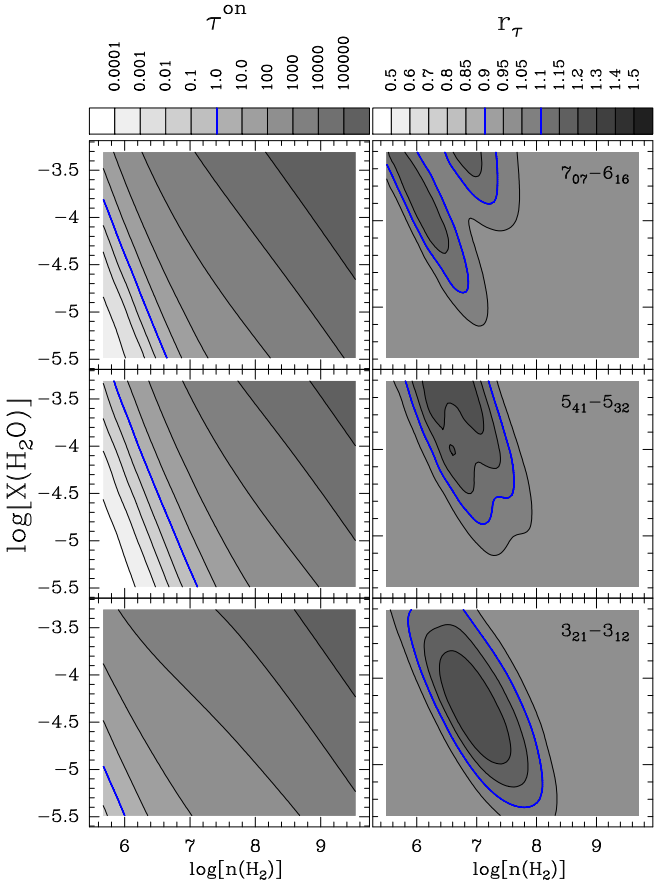}
\caption{The left column corresponds to the opacity of 
three purely thermal lines which are obtained in a treatment 
where masing lines are treated explicitly. The right 
column corresponds to the ratio $r_{\tau}= \tau^{off} / \tau^{on}$
 of the opacities obtained with the two treatments (see text). }
\label{app:thermiq}
\end{figure}

The two treatments for maser propagation differ in the evaluation 
of the average radiation field of the masing lines. Considering the 
current maser propagation treatment leads to higher values 
for the average radiation field which, in turn, when inverting 
the SEE, leads to a diminution in the population differences 
between levels involved in maser transitions. The modification 
of the populations of these levels subsequently modifies the 
whole energy level populations. This is illustrated in figure 
\ref{app:thermiq} for three lines which only show purely 
thermal emission. In this figure, we see that the opacities 
of these lines are modified and can show variations up to 
$\sim 25 \%$ between the two treatments in the region of 
the (n(H$_2$),X(H$_2$O)) plane where maser transitions 
are the most opaque. Nevertheless, for the range of parameters 
considered here, the number of lines which suffer from 
optical depth variations greater than 10\% 
is low, and for the majority 
of the thermal lines the opacities agree within a few 
percent between the two treatments. 

\begin{table*} 
\caption{Prediction of maser lines for different collisional rate coefficients sets.}
\begin{tabular}{lcccccc}
\hline
Transition & Frequency  & He & p-H$_2$ & p-H$_2$ & o-H$_2$ & o-H$_2$ \\
               & (GHz)  &  Green & SET1 & SET2 & SET1 & SET2  \\
\hline
$\left| \tau_{22} \right|$ &  & 8.2 (10.7) & 5.3 (7.1) & 9.1 (15.1) & 4.9 (6.8) & 7.6 (10.1) \\
\hline
$6_{1,6}-5_{2,3}$ & 22              		    & 1.00 (1.00) & 1.00 (1.00) & 1.00 (1.00) & 1.00 (1.00) & 1.00 (1.00) \\
$10_{2,9}-9_{3,6}$& 321            		   & 0.73 (0.88) & 1.11 (1.16) & 0.67 (0.70) & 1.17 (1.15) & 0.78 (0.89) \\

$4_{1,4}-3_{2,1}$& 380             		    & 0.53 (1.78) & 0.83 (2.14)& 0.50 (1.85) & 0.93 (2.35) & 0.59 (1.63) \\
$6_{4,3}-5_{5,0}$& 439           		            & 0.48 (0.63) & 0.83 (1.03)& 0.48 (0.69) & 0.91 (1.05) & 0.60 (0.93) \\
$7_{5,2}-6_{6,1}$& 443          		            & 0.02 (0.05) & 0.15 (0.21)& 0.11 (0.13) & 0.16 (0.24) & 0.17 (0.25) \\
$4_{2,3}-3_{3,0}$& 448          	        	            & 0.42 (1.49) & 0.68 (1.89)& 0.39 (1.69) & 0.74 (1.84) & 0.49 (1.55) \\
$1_{1,0}-1_{0,1}$& 557              		  & $<0.01$  & $<0.01$  & $<0.01$ & $<0.01$  & $<0.01$  \\
$5_{3,2}-4_{4,1}$& 621              	          & 0.45 (1.69) & 0.76 (2.31)& 0.42 (1.77) & 0.82 (2.13) & 0.53 (1.84) \\
$7_{2,5}-8_{1,8}$& 1146                                      & N (N) & $<0.01$ & N (N) &  $<0.01$  & N (N) \\
$\mathbf{3_{1,2}-2_{2,1}} $ & 1153                & 0.02 (0.02) & 0.17 (0.18) & $<0.01$ & 0.20 (0.26) & $< 0.01$ \\
$6_{3,4}-5_{4,1}$& 1158 		                   & 0.18 (0.80) & 0.36 (1.61) & 0.15 (0.56) & 0.44 (2.07) & 0.24 (1.37) \\
$\mathbf{8_{5,4}-7_{6,1}} $ & 1165               & 0.09 (0.11) & 0.21 (0.35) & 0.04 (0.04) & 0.21 (0.37) & 0.15 (0.24) \\
$7_{4,3}-6_{5,2}$& 1278 	         	           & 0.17 (0.49) & 0.38 (1.50) & 0.14 (0.41) & 0.43 (1.70) & 0.21 (0.79) \\
$8_{2,7}-7_{3,4}$& 1296 				   & 0.31 (0.76) & 0.54 (1.70) & 0.30 (0.90) & 0.59 (1.89) & 0.34 (0.78) \\
$8_{4,5}-9_{1,8}$& 1308				  & N (N) & N (N) & N (N) & N (N) & N (N) \\
$6_{2,5}-5_{3,2}$& 1322 				  & 0.19 (0.34) & 0.41 (0.95) & 0.21 (0.27) & 0.46 (0.99) & 0.26 (0.45) \\
$8_{5,4}-9_{2,7}$& 1596                            & N (N) & N (N) & N (N) & N (N) & N (N) \\
$8_{4,5}-7_{5,2}$& 1885 				   & 0.01 (0.02) & 0.06 (0.06) & 0.01 (0.01) & 0.10 (0.10) & 0.03 (0.03) \\
$5_{2,3}-4_{3,2}$& 1918 	 			   & $< 0.01$ & $< 0.01$ & $<0.01$ & 0.01 (0.01) & $< 0.01$ \\
$8_{3,6}-7_{4,3}$& 2245 			            & $< 0.01$ & 0.20 (0.30) & 0.01 (0.02) & 0.27 (0.43) & 0.01 (0.02) \\
$7_{3,4}-6_{4,3}$& 2567 		                   & $< 0.01$ & $< 0.01$ & N (N) & $< 0.01$  &  $< 0.01$ \\
$8_{3,6}-9_{0,9}$& 2577                             & N (N) & N (N) & (N) & N (N) & N (N) \\
$9_{2,7}-10_{1,10}$& 2619                         & N (N) & N (N) & (N) & N (N) & N (N) \\
$9_{4,5}-8_{5,4}$& 3150                             & N (N) & N (N) & (N) & N (N) & N (N) \\
\end{tabular}
\tablefoot{The first line gives the maximum 22 GHz opacity ($\tau_{22}$) obtained for the density and temperature range covered
in this study (see Sec. \ref{masers_H2O}). For each masing line, the ratio of the maximum opacity to $\tau_{22}$
is given, making use of the approximate treatment described in Sec. \ref{application} (numbers in parentheses) or with 
the current method.
The transitions in bold correspond 
to the maser transitions predicted by the present work and not predicted by \cite{yates1997}. The transitions
with opacity ratio below $10^{-5}$ as well as the transitions that do not show population inversion
are reported with the N symbol.}
\label{maser_liste1}
\end{table*} 

\begin{table*} 
\caption{Same as Table \ref{maser_liste1} but for p--H$_2$O}
\begin{tabular}{lcccccc}
\hline
Transition & Frequency  & He & p-H$_2$ & p-H$_2$ & o-H$_2$ & o-H$_2$ \\
               & (GHz) & Green &  SET1 & SET2  & SET1 & SET2  \\
\hline
$\left| \tau_{22} \right|$ & & 8.2 (10.7) & 5.3 (7.1) & 9.1 (15.1) & 4.9 (6.8) & 7.6 (10.1) \\
\hline
$3_{1,3}-2_{2,0}$ & 183              		     &  0.64 (1.22) &  0.98 (1.55) & 0.59 (1.29) &  1.08 (1.63) &  0.70 (1.16) \\
$5_{1,5}-4_{2,2}$ & 325             		     &  0.61 (1.24) &  0.85 (1.29) & 0.57 (1.35) &  0.93 (1.32) &  0.68 (1.22) \\
%$\mathit{10_{3,7}-11_{2,10}}$ & 390                        &  .  &  .   & . &  . &  . \\
$7_{5,3}-6_{6,0}$ & 437              		     &  $<0.01$ (0.02) &  0.16 (0.23) & 0.09 (0.12) &  0.15 (0.22) &  0.13 (0.21) \\
$6_{4,2}-5_{5,1}$ & 471              		     &  0.46 (0.65) &  0.80 (1.15) & 0.46 (0.70) &  0.83 (1.06) &  0.59 (0.97) \\
$5_{3,3}-4_{4,0}$ & 475              		     &  0.49 (1.40) &  0.77 (1.79) & 0.44 (1.42) &  0.86 (1.80) &  0.60 (1.60) \\
$6_{2,4}-7_{1,7}$ & 488              		     &  N (N) &  $<0.01$ & N (N) &  $< 0.01$ &  $< 0.01$ \\
%$\mathit{8_{6,2}-7_{7,1}}$ & 505                          &  .  &  .   & . &  . &  . \\
%$\mathit{9_{7,3}-8_{8,0}}$ & 646                          &  .  &  .   & . &  . &  . \\
$2_{1,1}-2_{0,2}$ & 752                       &  N (N)  &  N (N)   & N (N) &  N (N) &  N (N) \\
%$\mathit{11_{3,7}-12_{2,10}}$ & 767                      &  .  &  .   & . &  . &  . \\
$9_{2,8}-8_{3,5}$ & 906              		     &  0.45 (1.07) &  0.73 (1.41) & 0.43 (1.00) &  0.79 (1.32) &  0.49 (0.98) \\
$4_{2,2}-3_{3,1}$ & 916              		     &  0.25 (0.56) &  0.36 (0.84) & 0.25 (0.73) &  0.39 (0.92) &  0.29 (0.81) \\
$5_{2,4}-4_{3,1}$ & 970              		     &  0.26 (0.80) &  0.55 (2.77) & 0.21 (0.81) &  0.60 (3.06) &  0.32 (1.21) \\
$1_{1,1}-0_{0,0}$ & 1113                       &  N (N)  &  N (N)  & N (N) & N (N)  &  N (N) \\
$\mathbf{7_{4,4}-6_{5,1}}$ & 1173              		  &  0.19 (0.41) &  0.41 (0.96) & 0.18 (0.29) &  0.45 (0.98) &  0.23 (0.67) \\
$\mathbf{8_{5,3}-7_{6,2}}$ & 1191             		  &  0.18 (0.21) &  0.15 (0.24) & 0.14 (0.18) &  0.19 (0.29) &  0.24 (0.38) \\
%$\mathit{9_{6,4}-8_{7,1}}$ & 1215                       &  .  &  .   & . &  . &  . \\
%$\mathit{13_{3,11}-12_{4,8}}$ & 1272                       &  .  &  .   & . &  . &  . \\
$7_{4,4}-8_{1,7}$ & 1345                         &  N (N)  &  N (N)   & N (N) &  N (N) &  N (N) \\
$9_{4,6}-10_{1,9}$ & 1435              		  &  N (N) &  N (N) & N (N) &  N (N) &  N (N) \\
$7_{2,6}-6_{3,3}$ & 1441              		     &  0.25 (0.41) &  0.49 (1.61) & 0.23 (0.35) &  0.55 (1.88) &  0.29 (0.47) \\
$6_{3,3}-5_{4,2}$ & 1542              		     &  0.15 (0.15) &  0.28 (0.35) & 0.13 (0.11) &  0.32 (0.49) &  0.20 (0.35) \\
$7_{3,5}-6_{4,2}$ & 1766              		     &  0.06 (0.07) &  0.24 (0.39) & 0.04 (0.06) &  0.30 (0.50) &  0.07 (0.11) \\
$\mathbf{8_{4,4}-7_{5,3}}$ &  2162             		  &  $<0.01$ (N) &  $<0.01$ & $< 0.01$ &  $<0.01$ &  $< 0.01$ \\
$9_{3,7}-8_{4,4}$ & 2532              		  &  $< 0.01$ &  0.16 (0.18) & $<0.01$ (0.01) &  0.20 (0.23) &  $< 0.01$ \\
$\mathbf{9_{4,6}-8_{5,3}}$ &  2547         		   &  N (N) &  $<0.01$ & $<0.01$ &  $<0.01$ &  N (N) \\
$\mathbf{6_{2,4}-5_{3,3}}$ &  2962             		  &  N (N) &  $<0.01$ & N (N)&  $<0.01$ &  N (N) \\
$\mathbf{8_{3,5}-7_{4,4}}$ &  3670             		  &  N (N) &  $<0.01$ & N (N) &  $<0.01$ &  N (N) \\
\end{tabular}
\tablefoot{The ratios are still given as a function of the maximum
22 GHz opacity.}
\label{maser_liste2}
\end{table*} 

\section{Maser prediction} \label{masers_H2O}

Most of the information that can be extracted from water line observations 
relies on the modelling of its excitation. From this point of view, water is a difficult molecule
to treat since its high dipole moment causes most of its transitions to be sub--thermally 
excited \citep[see, e.g.][]{Cernicharo2006b}, harbouring very large opacities 
\citep[see, e.g.][]{Gonzalez98}, and many of them being maser in nature
\citep{Cheung69, Waters80, Phillips80, Menten90a, Menten90b, Menten91, Cernicharo1990, 
Cernicharo1994, Cernicharo1996, Cernicharo1999, Cernicharo2006b, Gonzalez95, Gonzalez98}. 
Accurate modeling thus requires, in addition to a good description of the source structure, 
the availability of collisional rate coefficients. Moreover, as discussed above,
special formalisms have to be developed in order to have a realistic treatment of the transfer of radiation
in the case of saturated masers.
The method developed in the present work for masing lines is used 
to assess which water lines are expected to show maser excitation 
as well as the relative opacity of the masers.  To do so, we have performed
calculations for various gas kinetic temperatures, H$_2$ and water volume densities.
The grid was set up using the boundaries  : $100 $ K $<$ T$_k$ $< 1000$ K, 
$10^6$ cm$^{-3} < $  n(H$_2$)  $ < 10^{10}$ cm$^{-3}$ and n(H$_2$O) 
$\in \left\{ 10^2 ; 10^3 ; 10^4 ; 10^5 \right\}$ cm$^{-3}$. 
Additionally, the parameter space is reduced by considering that the water abundance cannot exceed  a 
relative abundance of $10^{-4}$ with respect to H$_2$.
This range of values for the considered parameters is appropriate to the case
of the circumstellar envelopes of evolved stars.
The ratios of the line opacities with respect to the maximum 22 GHz opacity
are given in Table \ref{maser_liste1} for o--H$_2$O and 
Table \ref{maser_liste2} for p--H$_2$O. For the two water symmetries, we compare
the opacities obtained with the rate coefficients of \cite{faure2007} (indexed as SET1), 
\cite{dubernet2006,dubernet2009,daniel2010,daniel2011} (indexed as SET2) and \cite{green1993}.
In these tables, the ratios obtained with the approximation introduced by \cite{yates1997}
are indicated in parentheses.

The parameter space considered here is a sub--space of the parameter space
explored by \cite{yates1997}, since the maximum gas temperature in their 
calculation is 2000 K and because they include the pumping by dust photons.
In the present case we use a lower boundary for the temperature since, in order
to go higher, we would have to extrapolate the rate coefficients. Since we aim
to compare various sets of collisional rate coefficients, we choose not to introduce artefacts
due to the extrapolation procedure in the comparison of the results.
Dust can play an important role in deriving the populations of water levels. However,
since we report the ratio of the maximum opacity of the masing lines with respect to the 
maximum opacity of the 22 GHz line, the consideration of dust can only affect the current result
for the lines which are radiatively pumped. For these lines, the ratios reported in Tables  \ref{maser_liste1}
and \ref{maser_liste2} have to be considered as lower limits. On the other hand, the ratios reported should be accurate 
for the lines which are collisionally pumped.

In order to compare the current results with the results reported by \citet{yates1997}, we have
to convert the population density ratio of the latter study into opacity ratios. To do so, we start
from the definition of the opacity along a ray of length $L$ along which the density populations are uniform.
Assuming that the velocity profile $\phi(\nu)$ can be expressed as a gaussian, the opacity at the line $i$ centre is :
\begin{eqnarray}
\tau(i) 
& = & \frac{1}{8 \pi^{\frac{3}{2}}} \times \frac{c^3}{\nu_i^3} A(i)  \, g(i) \, \textrm{d}n(i) \,  \times L \times
\left[ \frac{2 k_B T_i}{m} + \sigma_{i}^2 \right]^{-\frac{1}{2}}
\label{comp_yates1}
\end{eqnarray}

where $\nu_i$ is the rest frequency of line $i$, $A(i)$ is the Einstein coefficient for spontaneous emission, $g(i)$ is the 
degeneracy of the upper level of the transition and d$n(i)$ is the population difference per sub--level. This latter quantity is the one 
considered by \citet{yates1997} and is defined as : 
\begin{eqnarray}
\textrm{d}n(i) = \left(\frac{n_l}{g_l} - \frac{n_u}{g_u} \right)
\end{eqnarray}
where the indexes $u$ and $l$ stand respectively for the upper and lower level of transition $i$. In Eq. \ref{comp_yates1},
the gas temperature $T_i$ and turbulence velocity $\sigma_i$ are indexed according to the line, since in the comparison, we only retain
the physical parameters that lead to the maximum opacity for this particular line. The opacity ratio thus correlate to the 
population density ratio considered by \citet{yates1997} through :
\begin{eqnarray}
\frac{\tau_{1}}{\tau_2} 
& = &   \left( \frac{\nu_2}{\nu_1} \right)^3 \frac{A(1)}{A(2)}  \, \frac{g(1)}{g(2)}  \frac{\textrm{d}n(1)}{\textrm{d}n(2)} \label{conversion} \\ & \times & 
\left[ \frac{2 k_B T_1}{m} + \sigma_1^2 \right]^{-\frac{1}{2}}
\left[ \frac{2 k_B T_2}{m} + \sigma_2^2 \right]^{\frac{1}{2}} \nonumber
\end{eqnarray}

Since the temperatures and turbulence of the models for which the ratios are given in Table 2 of \citet{yates1997}
are unknown, we assume $T_1 = T_2$ and $\sigma_1 = \sigma_2$ while
performing the conversion. The maximum error introduced by this assumption is typically of a factor  $\sim$ 2-3
(considering that the 22 GHz has its maximum inversion at $600$ K as stated in Yates (1997). )
By considering col. 5 of Table \ref{maser_liste3}, it can be seen 
that the values currently obtained and the one obtained by \citet{yates1997} can differ by factors 
larger than 2-3.
The differences might arise from the use of plane--parallel geometry in 
\cite{yates1997} against spherical geometry in the current work, or because we do not consider the effect 
introduced by the pumping by dust photons in the current study. By examining the results of the comparison 
reported for o--H$_2$O in Table \ref{maser_liste3} and for the lines with frequencies higher than 1.3 THz ($\lambda < 230$ $\mu$m), we can 
see that the ratios can take high values above this threshold frequency and tend to increase while the frequency increases. Since this behaviour
is correlated with the shape of the dust emission, it can be guessed that a quantitative comparison between the two works is plagued 
by the absence of consideration of dust radiation in the present work.

\begin{table*} 
\begin{center}
\caption{Comparison of the current results with
\citet{yates1997}'s results.}
\begin{tabular}{ccccc}
\hline
o/p--H$_2$O Transition & Frequency  (GHz)& Yates97 & current & Yates97/current \\
\hline
$6_{1,6}-5_{2,3}$ & 22                           & 1.0     & 1.00 & 1.00 \\
$10_{2,9}-9_{3,6}$& 321                        & 0.59   & 0.88 & 0.67 \\
$4_{1,4}-3_{2,1}$& 380                          & 1.46   & 1.78 & 0.82 \\
$6_{4,3}-5_{5,0}$& 439                          & 0.13   & 0.63 & 0.20 \\
$7_{5,2}-6_{6,1}$& 443                          & 0.14   & 0.05 & 2.89 \\
$4_{2,3}-3_{3,0}$& 448                          & 1.12   & 1.49 & 0.75 \\
$5_{3,2}-4_{4,1}$& 621                          & 0.36   & 1.69 & 0.21 \\
$6_{3,4}-5_{4,1}$& 1158                        & 0.89   & 0.80 & 1.11 \\
$7_{4,3}-6_{5,2}$& 1278                        & 0.09   & 0.49 & 0.19 \\
$8_{2,7}-7_{3,4}$& 1296                        & 1.74   & 0.76 & 2.29 \\
$6_{2,5}-5_{3,2}$& 1322                        & 3.11   & 0.34 & 9.17 \\
$8_{4,5}-7_{5,2}$& 1885                        & 0.31   & 0.02 & 15.7\\
$5_{2,3}-4_{3,2}$& 1918                        & 1.74   & $< 0.01$ & $>$ 173 \\
$8_{3,6}-7_{4,3}$& 2245                        & 0.86   & $< 0.01$ & $>$ 85   \\
$7_{3,4}-6_{4,3}$& 2567                        & 1.17   & $< 0.01$ & $>$ 116 \\ \hline
$3_{1,3}-2_{2,0}$ & 183   & 0.83   & 1.22 & 0.68 \\
$5_{1,5}-4_{2,2}$& 325  & 1.02   & 1.24 & 0.82 \\ 
$7_{5,3}-6_{6,0}$& 437    & 0.10   & 0.02 & 4.80 \\ 
$6_{4,2}-5_{5,1}$& 471    & 0.07   & 0.65 & 0.11 \\
$5_{3,3}-4_{4,0}$& 475    & 0.25   & 1.40 & 0.18 \\
$9_{2,8}-8_{3,5}$& 906    & 0.60   & 1.07 & 0.55 \\
$4_{2,2}-3_{3,1}$& 916    & 0.48   & 0.56 & 0.86 \\
$5_{2,4}-4_{3,1}$& 970    & 1.19   & 0.80 & 1.49 \\
$7_{2,6}-6_{3,3}$& 1441  & 1.07   & 0.41 & 2.62 \\  
$6_{3,3}-5_{4,2}$& 1542  & 0.11   & 0.15 & 0.72 \\
$7_{3,5}-6_{4,2}$& 1766  & 0.15   & 0.07 & 2.21 \\
\end{tabular}
\end{center}
\tablefoot{Comparison of the opacities obtained in the present work (column 4)
to the one estimated from \citet{yates1997} (column 3). In this latter case, the opacities
are obtained from the population density ratio they reported and making use of 
Eq. \ref{conversion} (see text for details). The last column gives the opacity ratio obtained 
in the two studies.}
\label{maser_liste3}
\end{table*}

By comparison to previous calculations \citep{neufeld1991,yates1997}, 
we find that the $8_{5,4}-7_{6,1}$ (1165 GHz), $7_{4,4}-6_{5,1}$ (1173 GHz),
$8_{5,3}-7_{6,2}$ (1191 GHz) lines can show population inversion with 
substantial opacities, independently of the set of rate coefficients used.
These lines were not predicted to be masers in the previous studies. 
The rest of the lines considered in this work agree with the predictions of \cite{yates1997}
with respect to the possibility of masing action, at least for the lines that show 
substantial inversion. 

Considering the effects introduced by the collisional rate coefficients, 
we find that the variations from one set to another are moderate. The lines which 
are found to be inverted are the same, independently of the set considered, and 
most of the opacities show variations of the order of a factor 2 or less.
Our method to treat the RT of the maser lines introduces large differences in 
the computed line opacities; indeed, for most of the lines, 
the opacities derived with the exact treatment are lower by up to a factor 4--5
in comparison to the opacities derived with the approximated treatment.
 
Interestingly, we find that many lines can show substantial inversion.
This is in agreement with the observations made by \cite{menten2008} who
report the fluxes of 9 maser transitions of water.
Note that the 437 GHz $7_{5,3}-6_{6,0}$ whose first detection is reported
in the latter study is expected to show a substantial inversion 
\citep[as was found in][]{yates1997} while this line was not found to be inverted
 by \cite{neufeld1991}.
 
 \begin{figure*}[h!]
\includegraphics[scale=0.70,angle = 270]{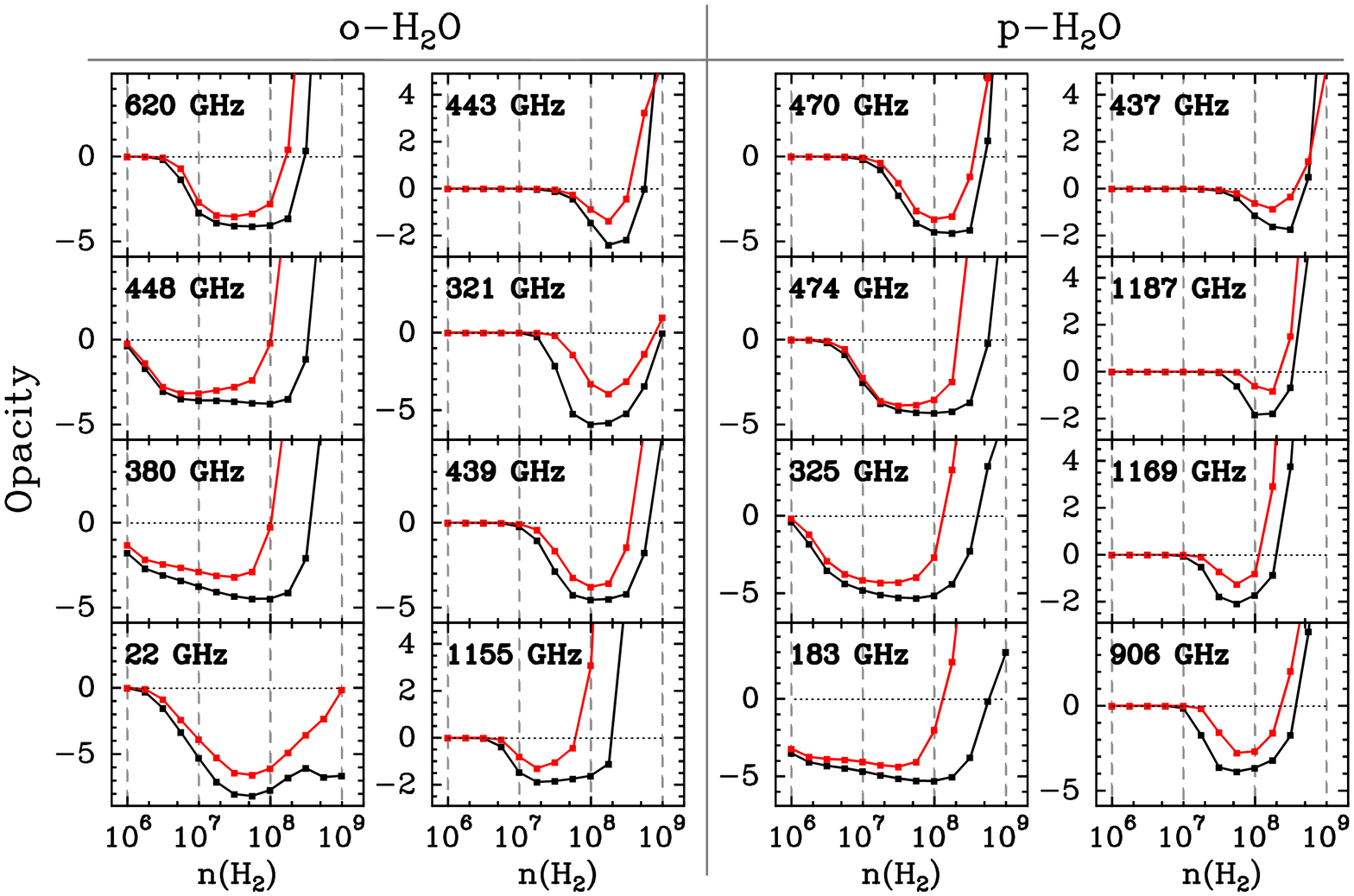}
\caption{Opacity as a function of H$_2$ volume density, for various o--H$_2$O and p--H$_2$O lines which are found to be substantially inverted at
T$_K$ = 500 K (red lines) and T$_K$ = 1000 K (black lines).  }
\label{zones_inversion}
\end{figure*}

Finally, \cite{humphreys2001} presented models for the excitation of the 22, 183, 321 
and 325 GHz water lines in the circumstellar envelopes (CSE) of AGB stars. They found 
the different masing lines to be excited in specific regions of the CSE, thus predicting different
morphologies of the emission maps for the masing lines they considered.
A direct comparison with their results is not possible, since in their study, the model that describes 
the circumstellar envelope is rather elaborated. Indeed, they first compute the H$_2$ volume density, gas temperature
and expansion velocity from an hydrodynamical code. As a second step,  they randomly define 
masing spots in the envelope and solve the 1D transfer problem for the various spots using the LVG 
approximation. The physical parameters that concern each emitting spot is defined according to its
distance to the star with respect to the parameters obtained from the hydrodynamical model.
Finally, the level populations derived from the LVG calculation are then used in a code that 
propagates the radiation and that 
takes into account maser saturation in order to calculate the emission of each spot.
Performing such a model is out of the scope of the current study, but it is possible to qualitatively
discuss their results with the current ones. In Fig. \ref{zones_inversion}, we show 
the opacities of various water maser lines as a function of the H$_2$ volume density. 
The rate coefficients used consider o--H$_2$ as a collisional partner and are taken from SET2.
The opacities are obtained using the treatment for maser propagation and 
for each line, the opacity is given at 500K and 1000K.

One of the conclusion drawn by 
\citet{humphreys2001} is that the 321 and 22 GHz are sensitive to the acceleration zone of the CSE.
Additionally, the 321 GHz line is predicted to trace only a region close to the star while the 
22 GHz emission extends further outside \citep[see Fig. 9 of ][]{humphreys2001}. Considering the 
opacities reported in Fig. \ref{zones_inversion}, we can see that the current results support those
findings. Indeed, at densities $\sim$ $10^9$ cm$^{-3}$, the 22 GHz line is the only masing line which 
is still subsequently inverted. At such a high H$_2$ density, the other masing lines are found to 
be collisionaly quenched. Moreover, the 321 GHz line is inverted only at relatively high H$_2$ densities, 
ie. n(H$_2$) $>$ $2 \, 10^7$ cm$^{-3}$ which support the fact that this line only traces the innermost 
part of the CSE.
\citet{humphreys2001} also find that the emission in the 325 GHz line resembles that of the 22 GHz.
In the current study, by examining the behaviour of the opacities of those two lines shown 
in Fig. \ref{zones_inversion}, we can indeed see that these lines have similar excitation conditions.
Additionaly \citet{humphreys2001} predicted the 183 GHz line to be inverted in the outermost part
of the CSE. In this region, the H$_2$ density is too low to produce inversion in the other masers.
Referring to Fig. \ref{zones_inversion}, we can see that at n(H$_2$) $\sim$ $10^6$ cm$^{-3}$, 
the 183 GHz line is indeed subsequently inverted while the other lines are not.
Finally, considering the behaviour of the opacities reported in Fig. \ref{zones_inversion}, 
it is possible to qualitatively predict the morphology that the lines not considered 
by \citet{humphreys2001} should have. As an example, the emission 
in the 380 GHz line should resemble the 183 GHz one and the 439 and 470 GHz lines should behave
similarly to the 321 GHz line.

\section{Conclusions}

The present work deals with obtaining a self--consistent 
solution for the problem of maser propagation within 
the scope of non--local radiative transfer. Test cases 
are presented for the water molecule in which the importance 
of dealing with maser propagation is emphasised, by 
comparison to a simplified treatment for the radiative 
transfer. The main conclusions are :

   \begin{enumerate}
      \item A self-consistent solution for the radiative 
transfer in masing lines can be obtained within the scope 
of the short characteristic method, with subsequent 
modification of the algorithm used to treat thermal lines 
due to the integro--differential nature of the transfer 
equation.
      \item The preconditioning of the statistical equilibrium 
equations is not feasible using the current method and the 
Ng acceleration technique cannot be used, so that the 
algorithm has the convergence rate of the Lambda iterative 
scheme.
      \item The test cases performed for the water 
molecule have shown that neglecting maser propagation can 
lead to substantial errors in the estimate of the masing 
line opacities, with differences greater than a factor 2. 
Dealing with strongly inverted masers, it means errors 
of several orders of magnitude in the prediction of line 
intensities.
      \item In the region where strong masing lines are 
present, the radiation field in these lines perturbs the 
whole population of the molecule. Therefore, including maser 
propagation also modifies the determination of line opacities 
for purely thermal lines and in the present test cases, 
maximum differences of the order of a few 10\% are found
between the two methods adopted. Nevertheless, the number 
of lines affected is rather low and most of the lines have 
identical optical depths in the two treatments. Given the 
larger computational time required to solve the maser propagation,
the approximation used in this study is a good 
alternative if one is concerned with interpreting the emission 
of non--masing lines.  
     \item We give predictions of H$_2$O maser opacities for physical 
conditions typical of circumstellar envelopes of AGB stars. In addition, 
the predictions based on various sets of rate coefficients involving 
He, p--H$_2$ or o--H$_2$ are compared. We find that from one set to another, 
the relative opacities of the masing lines are similar. However, the
absolute scale of the opacities are within a factor $\sim$ 2 
depending on the set used.   
    \end{enumerate}

\begin{acknowledgements}
The authors want to thank the referee M.D. Gray for its useful comments
that enable to improve the current manuscript. We also want to thank A. Baudry 
for useful discussions concerning the current study.
The authors want to thank J. Roberts 
and T. Bell for their careful reading of the manuscript.
This paper was partially supported
within the programme CONSOLIDER INGENIO 2010, under grant "Molecular
Astrophysics: The Herschel and ALMA Era.- ASTROMOL" (Ref.: CSD2009-
00038). We also thank the Spanish MICINN for funding support through
grants AYA2006-14876 and AYA2009-07304. 
\end{acknowledgements}

\end{document}